# *Is It Possible to OD on Alpha?*

Zura Kakushadze[§†¶1] and Jim Kyung-Soo Liew[‡‡2]

[§] *Quantigic® Solutions LLC,[3] 1127 High Ridge Road, #135, Stamford, CT 06905*

[†] *Department of Physics, University of Connecticut, 1 University Place, Stamford, CT 06901*

[¶] *Department of Theoretical Physics, A. Razmadze Mathematical Institute*
*I. Javakhishvili Tbilisi State University, 6 Tamarashvili Street, 0177 Tbilisi, Georgia*

[‡‡] *The Johns Hopkins Carey Business School, 100 International Drive, Baltimore, MD 21202*

Abstract

It is well known that combining multiple hedge fund alpha streams yields diversification benefits to the resultant portfolio. Additionally, crossing trades between different alpha streams reduces transaction costs. As the number of alpha streams increases, the relative turnover of the portfolio decreases as more trades are crossed. However, we argue, under reasonable assumptions, that as the number of alphas increases, the turnover does not decrease indefinitely; instead, the turnover approaches a non-vanishing limit related to the correlation structure of the portfolio's alphas. We also point out that, more generally, computational simplifications can arise when the number of alphas is large.

[1] Email: zura@quantigic.com

[2] Jim Kyung-Soo Liew, Ph.D., is an Assistant Professor in Finance at the Johns Hopkins Carey Business School. Email: kliew1@jhu.edu

[3] DISCLAIMER: This address is used by the corresponding author for no purpose other than to indicate his professional affiliation as is customary in publications. In particular, the contents of this paper are not intended as an investment, legal, tax or any other such advice, and in no way represent views of Quantigic® Solutions LLC, the website www.quantigic.com or any of their other affiliates.

## 0. Motivation and Summary

The quest to uncover, source, train, and ultimately combine hedge fund alpha streams appears never ending. At the end of 2013, the hedge fund industry stood with about $2.2 trillion dollars[4] and continues to grow each year. Many institutional investors have started to shy away from the hedge fund of funds vehicle in favor of plunging into direct hedge fund investments.[5] Additionally, some "bleeding-edge" institutional investors have gone as far as employing next-generation managed account platforms, which allow investors to pick and choose their hedge fund investments with the added comfort of knowing that position-level data exists. Given the continued financial innovation in the hedge fund industry, we add jet fuel to the fire by examining what happens when institutional investors have the ability to not only pick unique alpha streams (unique hedge fund strategies), but also the ability to "cross" trades between alpha streams.

Crossing trades is best described by a simple example. Consider two alpha streams generated by Manager A and Manager B. Suppose you gave $10 million to Manager A and you gave an equal allocation of $10 million to Manager B. The total amount invested would be $20 million. Over the course of time assume both Managers got their money invested. Now suppose Manager A wanted to sell a $1 million long position in, let's take for example, Telsa Motors Inc. (TSLA). Additionally, suppose that at that moment, Manager B wanted to buy a $1 million dollar investment in TSLA. If you could "cross" Manager A's sell orders with Manger B's buy orders, then you would just cross the $1 million of TSLA between Manager A and Manager B without going to the market, by giving Manager B the shares of TSLA that once belonged to Manager A in exchange for $1 million in cash.

The advantages of crossing trades are that both Manager A and Manager B save transaction costs. If there were no crossing of trades, Manager A would have sold $1 million of TSLA to the market and incurred the sales commission, (any) market impact, and paid – assuming no "smart" execution algorithms are employed – roughly half the bid-ask spread. Similarly, Manager B would have also incurred roughly half the bid-ask spread when buying $1 million of TSLA. As of 2:41:55 on March 24, 2014, TSLA was quoted with an ask price: $219.00 (×200) and a bid price: $218.73 (×100), on finance.yahoo.com, resulting in a $0.27 bid-ask

[4] As of Q4 2013 BarclayHedge.com estimated the hedge fund industry had over $2,156.7B AUM (Assets under management): http://www.barclayhedge.com/research/indices/ghs/mum/HF_Money_Under_Management.html

[5] There exists a rich literature on hedge funds; some notable work include: Ackermann, McEnally and Ravenscraft (1999), Fung and Hsieh (1999, 2000, 2001), Liang (1999, 2000, 2001), Agarwal and Naik (2000a, 2000b), Edwards and Caglayan (2001), Kao (2002), Brooks and Kat (2002), Amin and Kat (2003). Earlier works include Schneeweis, Spurgin and McCarthy (1996), Fung and Hsieh (1997a, 1997b), Brown, Goetzmann and Ibbotson (1999), Edwards and Liew (1999a, 1999b). Asness, Krail and Liew (2001) show the potential estimation problems with examination of hedge fund data. For an excellent introduction see Lo (2001) and Chan, Getmansky, Haas and Lo (2005).

spread (which is \$219.00 less \$218.73), or about 12 basis points (bps), which is the bid-ask spread divided by the average of the bid and ask quotes (the mid-quote) at that time.

Half this spread would have been 6 bps. Since each Manager A and Manager B incur roughly half the bid-ask spread when selling to/buying from the market, together they incur roughly the bid-ask spread. Crossing this trade, excluding sales commission and market impact, would have saved the investor 12 bps times \$1 million or \$1,200. Not that much some would say, but consider this crossing could have happened across a large portfolio with over \$1 billion invested in 100 unique alpha streams. Also, if we assumed that such a crossing trade occurred for each alpha stream every day (and in some cases multiple times intraday for different securities), the savings become substantial. Consider that in 1 year the investor would have incurred the bid-ask spread cost alone of: (252 days) $\times$ (100 alpha streams) $\times$ (\$1 million crossed amount / alpha stream) $\times$ (12 bps cost / crossed amount) = \$30.24 million dollars per year that could have been roughly saved by internally crossing such trades under our assumptions. Another way to look at these savings is that \$30.24 million / \$1 billion = 3.024%, which is roughly double of the fixed management fee of 1-2% per year typically charged by hedge funds. Savings from crossing are in the ballpark of the fixed management fees.

In this note, we ponder the future of assembling alpha streams from a perspective of a large institutional investor. We assume that many hedge fund alphas are readily available to us on such a platform and we can easily combine these distinct alpha streams. Moreover, we assume that these platforms give us the ability to "cross" trades between distinct alpha streams.

Under such a framework, we examine the advantages of "crossing" trades, that is, matching buy orders and sell orders across distinct strategies. As more and more alpha streams are combined, the possibility of crossing should increase, and therefore the percentage of the dollar turnover with respect to the total dollar investment – which percentage we refer to simply as "turnover" – should decrease. With certain reasonable assumptions we argue that turnover indeed decreases and converges to a non-vanishing limit. In fact, the notorious correlation between alpha streams plays a chief role in this limit.

To summarize, a few practical implication results from our theoretical exercise: First, institutional investors should absolutely argue for the ability to "cross" trades on these hedge fund platforms. Second, investors should keep their focus on combining less correlated alpha streams. Third, adding more and more alphas streams reduces portfolio turnover only if the new alpha streams are less correlated with the old alphas streams.

This paper is organized as follows: definitions are in Section 1, Sections 2 and 3 provide mathematical details, Section 4 discusses caveats, and Section 5 sets forth the "Large $N$ Limit".

## 1. Definitions

We begin with a set of alphas $\alpha_i, i = 1, \dots, N$, which are combined with weights $w_i > 0$:

$$\alpha \equiv \sum_{i=1}^{N} w_i \alpha_i \qquad (1)$$

$$\sum_{i=1}^{N} w_i = 1 \qquad (2)$$

Profit and Loss (P&L) $P \equiv \alpha I$, where $I$ is the investment level; for example, we set $I$ equal to $1 billion. For definiteness, let $P$ be daily P&L, so alphas are daily returns, albeit the actual horizon is immaterial for the following discussion.

For each $\alpha_i$ "turnover" $T_i \equiv D_i \,/\, I_i$ , where $D_i$ is the (daily) dollar amount traded if the amount $I_i$ is invested into $\alpha_i$. Notice that turnover is measured in percentage terms, not in dollars. If each of our 100 alpha streams had $10 million invested and each traded $1 million per day, the daily turnover of each alpha would be $1M / $10M = 0.1.

In this note our calculations are carried out in the framework where each alpha is traded in its own separate aggregation unit, and matching trades are crossed between separate aggregation units. If we did the same calculation assuming that all alphas are traded in a single aggregation unit, then the total investment $I$ would be reduced (compared with the separate aggregation units framework) by some factor – call it $\zeta$ – due to some alphas having desired holdings with opposite signs, which sometimes is referred to as "netting". However, the total dollar turnover of the combined alphas is unchanged (compared with the separate aggregation units framework), so what we refer to as turnover (*i.e.*, the percentage of the dollar turnover with respect to the total dollar investment) is enhanced by the same factor $\zeta$. We will comment on how to compute this factor in the single aggregation unit framework in Section 5.

Here, our goal is to understand how the turnover behaves with the number of alphas $N$ in the separate aggregation units framework.

## 2. Combining Two Alphas

We will approach the problem of understanding how the turnover $T$ behaves with $N$ by understanding the turnover of 2 combined alphas. Let's continue our example with Manager A ($10 million) and Manager B ($10 million). Note that we can extend this to combine 4 alphas by first combining 2 alphas into one alpha, 2 other alphas into another alpha, and then combining the resulting 2 alphas. By repeating this process, we can combine $N = 2^k$ alphas and see how $T$ behaves with $N$.

So, let us combine 2 alphas, $\alpha_1$from Manager A and $\alpha_2$ from Manager B, with weights $w_1$ and $w_2$ ($w_1 + w_2 = 1$) and turnovers $T_1$ and $T_2$, respectively. To simplify things, we will ignore trading bounds (*i.e.*, any restrictions on how much dollar amount can be traded for each stock). If there is no internal crossing between the trades generated by $\alpha_1$ and $\alpha_2$, then the turnover of the combined alpha is simply $T_* = w_1 T_1 + w_2 T_2$. Internal crossing reduces $T_*$.

Let us parameterize the internal crossing as follows. Let $D_1$be the total dollar amount traded by $\alpha_1$and $D_2$ be the total dollar amount traded by $\alpha_2$. Then $D_1 = T_1 I_1 = w_1 T_1 I$, and $D_2 = T_2 I_2 = w_2 T_2 I$, where $I_1 \equiv w_1 I$ and $I_2 \equiv w_2 I$ are the investment levels allocated to $\alpha_1$and $\alpha_2$from the total investment level $I$ (recall that $w_1 + w_2 = 1$). Let $\Delta$ be the dollar amount traded by $\alpha_1$and $\alpha_2$ in the opposite directions, *i.e.*, the crossed amount. Note that $0 \leq \Delta \leq \Delta_{max} \equiv \min(D_1, D_2)$. Let $\xi \equiv \Delta/\Delta_{max}$, so that $0 \leq \xi \leq 1$. Here $\xi = 0$ corresponds to no internal crossing, and $\xi = 1$ corresponds to 100% internal crossing.

With the above definitions, we have:

$$T_* = (1 - \xi)(w_1 T_1 + w_2 T_2) + \xi\, |w_1 T_1 - w_2 T_2| \qquad (3)$$

Let $\rho \equiv 1 - 2\xi$. Then

$$T_* = \frac{1+\rho}{2}(w_1 T_1 + w_2 T_2) + \frac{1-\rho}{2}|w_1 T_1 - w_2 T_2| \qquad (4)$$

When $\rho = 1$ (no internal crossing), we have $T_* = w_1 T_1 + w_2 T_2$, and when $\rho = -1$ (100% internal crossing), we have $T_* = |w_1 T_1 - w_2 T_2|$.

Let us illustrate the above discussion on our example with Manager A and Manager B. Table I gives an example where Manager A buys \$1M and sells \$1M and Manager B buys \$800K and sells \$800K worth of different combinations of stocks, so we have $D_1 =$ \$2M and $D_2 =$ \$1.6M. As before, let us assume that each Manager has a \$10M investment level, so $I_1 = I_2 =$ \$10M and $I =$ \$20M. Then Manager A has the turnover $T_1 =$ \$2M/\$10M = 0.2, and Manager B has the turnover $T_2 =$ \$1.6M/\$10M = 0.16. In this case we have $\Delta =$ \$700K worth of trades overlapping in opposite directions (\$200K in MSFT and \$500K in AAPL), $\Delta_{max} =$ \$1.6M, and $\xi = 0.4375$ and $\rho = 0.125$. If the overlapping buy and sell trades are crossed between Manager A and Manager B, the resulting net buy and sell amounts are summarized in the rows of Table I labeled "Net buys" and "Net sells", respectively. The buys total \$1.1M and the sells total \$1.1M. The total investment level is \$20M, so the turnover is $T_* =$ \$2.2M/\$20M = 0.11. And this is exactly what we get from Eq. (4) with equal weights $w_1 = w_2 = 1/2$ and $\rho = 0.125$.

Thus, $\rho$ parameterizes the overlap between long and short trades generated by the 2 alphas, *i.e.*, internal crossing. It is not necessarily the same as the correlation between the 2 alphas. However, there is a relation between the two. To simplify things, in the following we will *assume* that $\rho$ is simply the correlation between the 2 alphas. In Section 4 we elaborate on this simplifying assumption and argue that it is actually not critical to our main result when the number of alphas $N$ is large.

Furthermore, for equal weights ($w_1 = w_2 = 1/2$) and turnovers ($T_1 = T_2 \equiv \tau$) we have

$$T_* = \frac{1+\rho}{2}\,\tau \qquad (5)$$

To simplify things, in the following we will assume equal weights ($w_i = 1/N,\ i = 1, \dots, N$) and uniform turnovers ($T_i \equiv T[0] \equiv \tau$). We will also assume that the correlation of each alpha $\alpha_i$ with another alpha $\alpha_i$ $(i \neq j)$ is $\rho \equiv \rho[0] > 0$ (see Section 4).

To continue our example, we have assumed 100 alpha streams, so $w_i = 1/100,\ i = 1, \dots, 100$, and in Section 1 we assumed uniform turnover of \$1 million for each manager so $\tau =$ \$1M/\$10M $= 0.1$.

## 3. Iterative Procedure

Here is a descriptive summary of our iterative procedure below. Let's assume that the number of alphas $N$ is a power of 2. Let's further assume that all alphas have the same pair-wise correlation $\rho$. Let's take a set of any $2^{k-1}$ alphas with equal weights and call it Set-1. Then let's take a different (non-overlapping with Set-1) set of $2^{k-1}$ alphas with equal weights and call it Set-2. Then $\rho[k]$ is the correlation between Set-1 and Set-2. We can apply the results of the Appendix, namely, Eq. (23), to Set-1 and Set-2 with $\rho[k]$ identified with $\eta$ and $\rho[k-1]$ identified with $\gamma$. Then Eq. (7) below is simply Eq. (23) of the Appendix. Furthermore, Eq. (6) below is just Eq. (5) applied to Set-1 and Set-2. We then have an iterative procedure whereby we start with 2 alphas, combine them, then combine the resulting combination of 2 alphas with another combination of 2 alphas, then combine this combination of 4 alphas with another combination of 4 alphas, *etc*., so schematically we have 1 + 1 = 2, 2 + 2 = 4, 4 + 4 = 8, 8 + 8 = 16,…, $2^{k-1}$ + $2^{k-1}$ = $2^k$,… At each of these steps we apply Eq. (23) of the Appendix and Eq. (5) of Section 2. Then we sew everything neatly together into the final result. Here is the actual iterative procedure.

Next, we will combine $2^{k+1}$ alphas iteratively by combining a pair of $2^k$ alphas. Based on Eq. (5), the resulting turnover for $2^{k+1}$ alphas is given by:

$$T[k+1] = \frac{1+\rho[k]}{2}\,T[k] = \frac{T[0]}{2^{k+1}}\prod_{l=0}^{k}(1+\rho[l]) \qquad (6)$$

where $T[k]$ is the turnover of $2^k$ alphas, and $\rho[k]$ $(k > 0)$ is the correlation between 2 alphas each corresponding to combining $2^{k-1}$ alphas. We have (see the Appendix)

$$\rho[k] = \frac{2\rho[k-1]}{1+\rho[k-1]} \qquad (7)$$

This gives:

$$\rho[k] = \frac{2^k\rho[0]}{1+(2^k-1)\rho[0]} \qquad (8)$$

$$1+\rho[k] = \frac{1+(2^{k+1}-1)\rho[0]}{1+(2^k-1)\rho[0]} \qquad (9)$$

$$\prod_{l=0}^{k}(1+\rho[l]) = 1+(2^{k+1}-1)\rho[0] \qquad (10)$$

$$T[k+1] = \frac{T[0]}{2^{k+1}}\,(1+(2^{k+1}-1)\rho[0]) \qquad (11)$$

Identifying $2^{k+1}$ with $N$, for the turnover $T$ of $N$ alphas, we have

$$T = \tau\left[\rho + \frac{1-\rho}{N}\right] \qquad (12)$$

So, when $N$ is large, the turnover $T$ does not decrease to zero with increasing $N$ but approaches

$$T_{limit} \equiv \tau\rho \qquad (13)$$

This is because when more and more alphas are added, the correlation $\rho[k]$ between any two subgroups with (approximately) equal numbers of alphas approaches 1, as can be seen from Eq. (8), *i.e.*, such subgroups become more and more correlated with each other.

## 4. The Upshot and Comments

The above results indicate that when the number of alphas is large, adding more alphas reduces turnover only if the new alphas are less correlated with the old alphas (and each other) as compared with the correlations between the old alphas. If new alphas are not less correlated, then the turnover is not reduced.

There are some simplifying assumptions that went into the above discussion. However, they do not modify the above conclusion. For the sake of simplicity we assumed that the alpha weights $w_i$ and the individual alpha turnovers $T_i$ are uniform. Non-uniformity affects (4), (5), (12) and (13) in two ways: 1) it replaces $\tau$ in (12) and (13) with the weighted average

$$\tau = \sum_{i=1}^{N} w_i T_i \qquad (14)$$

and 2) it introduces the term proportional to $1 - \rho$ in (4). However, the latter does not change the above conclusion as in the large $N$ limit $\rho[k] \to 1$, so when $N$ is large we still have (13) with $\tau$ given by Eq. (14).

Another assumption is that $\rho$ is a correlation. As mentioned in Section 2, $\rho$ actually parameterizes internal crossing. Clearly, there is a relation between $\rho$ and the correlation, but in real life they need not be the same. For example, consider two alphas, $\alpha_1$ and $\alpha_2$, where $\alpha_1$ trades stocks A and B, and $\alpha_2$ trades stocks B and C. The correlation between $\alpha_1$ and $\alpha_2$ depends on the correlations between relevant returns of A, B and C and their variances. However, even if $\rho$ is not the same as the correlation, the large $N$ bound (13) still exists assuming the total number of stocks traded is bounded. Indeed, we know that in the large $N$ limit the correlation $\rho[k] \to 1$, *which means that internal crossing cannot grow indefinitely with* $N$, or else the correlation $\rho[k]$ would not approach 1. Again, while $\rho$ may not be exactly the same as the correlation, the two are related and the correlation can be used in lieu of the internal crossing parameterization to *estimate* the bound (13). Note that large $N$ is the key.

Above we also assumed that $\rho$ is uniform, *i.e.*, the correlations between all $\alpha_i$ and $\alpha_j$ ($i \neq j$) are the same. In reality these correlations $\rho_{ij}$ are non-uniform. However, $\rho$ can be thought of as appropriately averaged $\rho_{ij}$. One can be more precise and break $\rho_{ij}$ into appropriately defined quantiles, discard some number of the lowest and highest quantiles, and use $\rho_{lower}$ and $\rho_{upper}$ to constrain $T_{limit}$ in Eq. (13):

$$\tau\rho_{lower} \lesssim T_{limit} \lesssim \tau\rho_{upper} \qquad (15)$$

This assumes $\rho_{lower} > 0$.

In general, the above discussion assumes that $\rho > 0$. This is because one cannot have a large number of alphas all correlated with each other with correlation $\rho < 0$. For example, in the case of 4 alphas discussed in the Appendix, the lower bound on $\gamma$ is $-1/\,3$. In general, the lower bound on $\rho$ is $-1/(2^k - 1)$, which approaches 0 in the large $N$ limit. In the case of non-uniform $\rho_{ij}$, some $\rho_{ij}$ can be negative, but in the large $N$ limit, with the appropriately defined and truncated quantiles, $\rho_{lower}$ is positive. For an *illustrative* example see Fig.1.

## 5. The Power of the Large $N$ Limit

Above we made certain assumptions, including that $\rho$ is a correlation, and that all correlations are uniform, yet we argued that in the large $N$ limit none of these assumptions would modify our main result, which is that in this limit the turnover does not vanish but is expected to approach a finite limit. This is an example of the power of the large $N$ limit.

More generally, in the large $N$ limit some (albeit not necessarily all) quantities exhibit dramatic simplifications. A typical such quantity, call it $A$, will have an expansion

$$A = A_0 + \frac{A_1}{N} + \frac{A_2}{N^2} + \cdots + \frac{A_k}{N^k} + \cdots \qquad (16)$$

in a power series[6] of $1/N$. In the large $N$ limit the leading term $A_0$ survives and the sub-leading terms vanish. In many cases it is possible to get a handle on $A_0$ analytically, while the sub-leading terms may be much harder to compute. When we have portfolios with large numbers of alphas, the leading term in some cases might be all we need. In such cases the large $N$ limit provides a powerful tool, both computationally and to get conceptual insight.

A large $N$ limit has been extensively used in theoretical physics for four decades – see 't Hooft (1974).[7] Quantitative finance apparently has already entered an era where the large $N$ limit now becomes not only relevant but a reality with ever-increasing numbers of alpha streams. We foresee the large $N$ limit being used more and more extensively. In fact, using the power of the large $N$ limit it is possible to model the turnover reduction in the case of a general correlation matrix with non-uniform correlations, which will be reported shortly in a forthcoming publication.

Finally, let us comment on the investment level reduction (or "netting") in the framework where all alphas are traded in the same aggregation unit – see the end of Section 1. This effect can be modeled in the same way we modeled turnover reduction in Section 3. In fact, the formulas are all the same with turnover replaced by the investment level and with $\rho$ replaced by a parameter, call it $\psi$, that parameterizes "netting" ($\psi = 1$ corresponds to zero "netting", $\psi = -1$ corresponds to 100% "netting"). This $\psi$, just like $\rho$, is not necessarily the same as the correlation between the alphas, but it is related to it. In the large $N$ limit one can use the correlation in lieu of $\psi$ (just as we used it in lieu of $\rho$) to estimate the "netting" effect. In the large $N$ limit the "netting", just like turnover, has a bound, *i.e.*, the investment level doesn't get reduced indefinitely but is bounded, just as turnover is.

---

[6] In some cases there can be logarithmic corrections, *e.g.*, $\sim \ln(N)/N$.

[7] In theoretical physics the large $N$ limit was applied to the quest to solve Quantum Chromodynamics, a highly nonlinear theory that describes quarks and gluons, the building blocks of nuclei in atoms.

## Appendix: Correlations

Let $X_A, A = 1,2,3,4$ be random variables with 0 means and unit variances with correlation between each pair $X_A$ and $X_B$ ($A \neq B$) equal $\gamma$ (here $\langle *,* \rangle$ stands for a covariance):

$$\langle X_A, X_A \rangle = 1 \qquad (17)$$

$$\langle X_A, X_B \rangle = \gamma, \quad A \neq B \qquad (18)$$

Let

$$Y_1 = \frac{1}{2}(X_1 + X_2) \qquad (19)$$

$$Y_2 = \frac{1}{2}(X_3 + X_4) \qquad (20)$$

Then

$$\langle Y_1, Y_1 \rangle = \langle Y_2, Y_2 \rangle = \frac{1}{2}(1+\gamma) \qquad (21)$$

$$\langle Y_1, Y_2 \rangle = \gamma \qquad (22)$$

$$\eta \equiv \frac{\langle Y_1, Y_2 \rangle}{\sqrt{\langle Y_1, Y_1 \rangle}\sqrt{\langle Y_2, Y_2 \rangle}} = \frac{2\gamma}{(1+\gamma)} \qquad (23)$$

Here $\eta$ is the correlation between $Y_1$ and $Y_2$.

## Tables

| | MSFT | IBM | AAPL | DELL | ORCL | JAVA |
|---|---|---|---|---|---|---|
| $\alpha_1$ buys | $200K | $300K | --- | --- | $500K | --- |
| $\alpha_1$ sells | --- | --- | $600K | $400K | --- | --- |
| $\alpha_2$ buys | --- | --- | $500K | --- | $300K | --- |
| $\alpha_2$ sells | $400K | --- | --- | $200K | --- | $200K |
| Net buys | --- | $300K | --- | --- | $800K | --- |
| Net sells | $200K | --- | $100K | $600K | --- | $200K |
| $\Delta$ | $200K | --- | $500K | --- | --- | --- |

Table I. Here $\alpha_1$ and $\alpha_2$ stand for Manager A and Manager B, respectively. We show dollar amounts bought or sold by each Manager. We have arranged the numbers such that Manager A buys $1M and sells $1M worth of stock, and Manager B buys $800K and sells $800K worth of stock. We also show net buys and sells after crossing trades between Manager A and Manager B. In the last row we give contributions into $\Delta$ from individual stocks.

## Figures

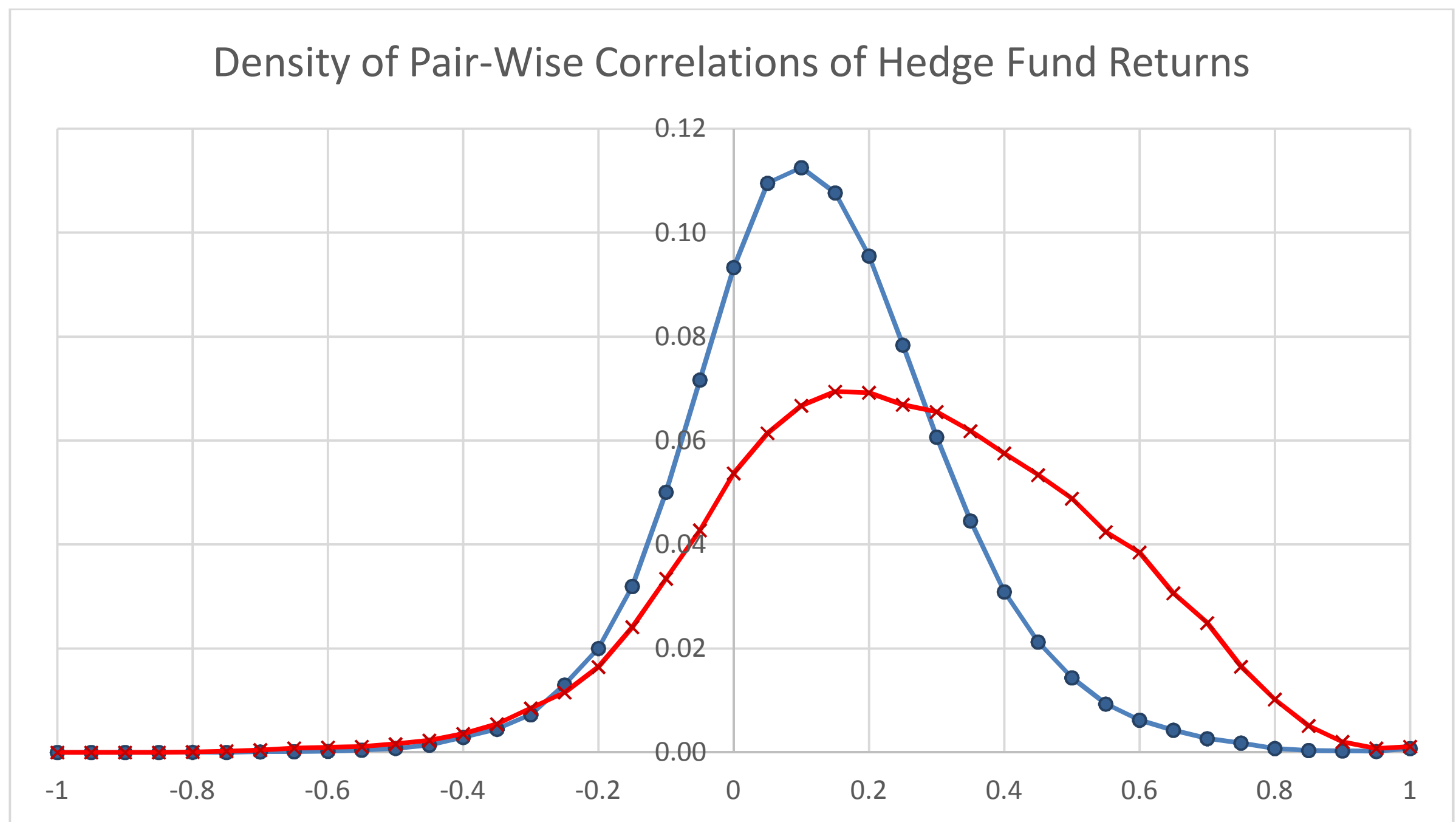

Fig.1. Density of off-diagonal correlations of 657 hedge fund returns ("HF") from Morningstar data for 1990-2014 based on i) raw HF (red line) and ii) HF adjusted for RF (whose effect is small) and regressed over Mkt-RF and Fama-French risk factors SMB, HML, WML (blue line) – we add the intercepts (these have no effect) to the residuals and compute the correlations.